\documentclass{aa}  
\usepackage{graphicx}
\usepackage{txfonts}
\usepackage[figuresright]{rotating}
\begin{document}
\title{Optical and infrared properties of V1647 Orionis during the 2003-2006 outburst}
   \subtitle{II. Temporal evolution of the eruptive source}
\author{D. Fedele \inst{1,2}, M. E. van den Ancker\inst{1}, M. G. Petr-Gotzens\inst{1} \and P. Rafanelli\inst{2}
\fnmsep\thanks{Based on observations collected at the European Southern Observatory, Paranal, Chile. Proposal ID: 272.C-5045, 074.C-0679, 276.C-5022}}

\offprints{D. Fedele}

\institute{European Southern Observatory, Karl Schwarzschild Strasse 2, D-85748 Garching bei M\"unchen, Germany\\
\email{dfedele@eso.org}
\and
Dipartimento di Astronomia, Universit\'a degli studi di Padova, Vicolo dell'Osservatorio 2, 35122 Padova, Italy}

%\date{DRAFT,Apr 25 2007}
%
\authorrunning{D. Fedele et al.}
\titlerunning{Temporal evolution of the outburst of V1647 Ori}
\abstract
{}
{The occurrence of new FU Orionis-like objects is fundamental to understand the outburst mechanism in young stars and their role in star
formation and disk evolution. Our work is aimed at investigating the properties of the recent outburst of V1647 Ori.
}
{Using optical and mid infrared long slit spectroscopy we monitored V1647 Ori in outburst between February 2004 and January 2006.
}
{The optical spectrum is characterized by H$\alpha$ and H$\beta$ in P-Cygni profile and by many weak \ion{Fe}{I} and \ion{Fe}{II} 
emission lines. Short timescale variability was measured in the continuum and line emission. On January 2006 we detected for the first 
time forbidden emission lines ([\ion{O}{I}], [\ion{S}{II}] and [\ion{Fe}{II}]). These lines are likely produced by an Herbig-Haro object
driven by V1647 Ori. The mid infrared the spectrum of V1647 Ori is flat and featureless at all epochs. The SED changed drastically: the 
source was much redder in the early outburst than in the final phase. The magnitude rise and the SED of V1647 Ori resembles that of a 
FUor while the duration and recurrence of the outburst resemble that of a EXor. The optical spectrum is clearly distinct from either the
absorption line spectrum of a FUor or the T Tauri-like spectrum of an EXor.} 
{Our data are consistent with a disk instability event which led to an increase of the mass accretion rate. The data also suggest the 
presence of a circumstellar envelope around the star+disk system. The peculiar $N$ band spectrum might be explained by dust sublimation 
in the outer layers of the disk. The presence of the envelope and the outburst statistics suggest that these instability events 
occur only in a specific stage of a Class I source (e.g. in the transition phase to an optically visible star surrounded by a 
protoplanetary disk). We discuss the outburst mechanisms in term of the thermal instability model.
}
\maketitle
\section{Introduction}
One of the clearest pieces of evidence for disk accretion during early stages of stellar evolution are events like FU Orionis and EX Lupi
outbursts (Herbig \cite{herbig77}, Hartmann \& Kenyon \cite{hk}). 
These outbursts are thought to be the consequence of a sudden and steep increase of the mass accretion rate onto the central star, which
changes from those commonly found around T Tauri stars ($\sim$ 10$^{-7}$ M$_\odot$~yr$^{-1}$) into values of 10$^{-3}$--10$^{-4}$ 
M$_\odot$~yr$^{-1}$. Statistical studies suggest that young low-mass stars experience several FU Orionis like outbursts during the early 
phase of stellar evolution. The emergence of a new pre-main-sequence outburst object is then a unique opportunity to address the 
physical process that occurs in the disk's interior. For this reason several astronomers have focused their attention on the recent 
outburst of V1647 Ori.
\noindent

V1647 Ori is a young eruptive star known to be the illuminating source of McNeil's Nebula, a reflecting nebula discovered by the 
amateur astronomer J.W. McNeil in January 2004 (McNeil et al. \cite{mcneil}). In the months following the discovery, the star, located 
in the L1630 cloud within the Orion B molecular cloud complex, showed an increase of its optical/IR brightness of up to 6 magnitudes. 
The outburst has been observed from the X-ray regime (e.g. Grosso et al. \cite{grosso}) to infrared wavelength (see e.g. Muzerolle et al.
\cite{muzerolle}, Andrews et al. \cite{andrews}). In February 2004, 4 months after the onset of the outburst, the brightness rise 
stopped and the  magnitude remained (relatively) constant. In November 2005, a communication by Kospal et al. (\cite{kospal}) claimed the
beginning of a fast fading phase in the optical light of V1647 Ori. The system is further characterized by a red energy 
distribution and by many emission lines in its optical and near-IR spectrum. Apart from the Brackett series (seen purely in emission), 
all the Hydrogen lines exhibit a 
P-Cygni profile, which indicates mass outflow in a wind. Vacca et al. (\cite{vacca}) find that their near-IR emission line spectrum is 
consistent with a dense and ionized wind model where the optically thick H lines are produced. The same model is able to explain why the 
outburst has not been seen at radio wavelength (Andrews et al. \cite{andrews}). V1647 Ori is known to have experienced a previous 
outburst as is clear from the appearance of the reflection nebula in 1996--1997 in the atlas of \cite{mallas} and as recently confirmed 
by Aspin et al. (\cite{aspin}). Furthermore, its optical and near-IR spectrum does not resemble any other previous 
spectra of FUors or EXors objects (Reiburth \& Aspin \cite{reipurth}, Vacca et al. \cite{vacca}). The 2-3 years duration of the outburst,
its recurrence on a timescale of decades and the ``peculiar'' spectrum of V1647 Ori, are important clues for the comprehension of 
outburst events in pre-main-sequence stars.
\noindent

From February 2004 to January 2006 we have followed the evolution of the outburst of V1647 Ori at optical (4700-7300 \AA) and mid-IR 
(8-13 $\mu$m) wavelengths. Here, we present the results of our photometric and spectroscopic monitoring of the eruptive source. The 
analysis of the reflection nebula are presented in a separate paper (Fedele et al. \cite{fedele}, hereafter paper I). Observations and 
data reduction are described in section \ref{sec:obs}. In section \ref{sec:results} we analyze the observations. A discussion of the 
results is presented in section \ref{sec:discussion}. We draw the conclusions in \ref{sec:conclusions}.
\section{Observations and data reduction}\label{sec:obs}
Observations were performed using FORS2 at ESO's Very Large Telescope in Paranal, Chile and TIMMI2 at the 3.6 m telescope at La Silla.
FORS2 (\cite{appenzeller}) is an optical facility (3000-10000 \AA) which allows imaging in different bands and grism spectroscopy.
TIMMI2 (\cite{kaeufl}) is a mid-infrared (8-14 $\mu$m) multi-mode instrument including low- and medium-resolution spectrograph. We have 
also included in our analysis some publicly available VLTI/MIDI observations of V1647 Ori taken from the ESO 
archive\footnote{{\it http://www.eso.org/archive}}. MIDI is the mid-infrared beam-combiner facilities of the ESO VLT interferometer 
(Leinert et al. \cite{leinert}). 
\subsection{Optical Spectroscopy}
12 long slit spectra of V1647 Ori were obtained with FORS2 and with the grism 1400V (4560-5860\AA, $\lambda / \Delta \lambda$ $\sim$ 
2100) between 2004 February 18 and 2005 December 27. Seven further long slit spectra were obtained between 2004 December 08 and 2006 
January 29, with the grism 1200R (5750-7310\AA, $\lambda / \Delta \lambda$ $\sim$ 2100). A detailed log of the observations is reported 
in Table \ref{tab:speclog}.
\noindent  

A standard optical long slit spectra extraction procedure was applied to reduce the raw data - bias subtractions, flat-fielding, 
wavelength calibration, cosmic rays and sky background removal and weighted average along the spatial axis. Observations of 
spectro-photometric standard stars during each night allowed us to compute the sensitivity function of the spectrograph. To 
flux-calibrate the spectra of V1647 Ori, the 1-dimensional extracted spectra were first multiplied by the sensitivity function -- to 
compute the exact slope of the spectrum -- and then scaled to the flux level measured from the acquisition images. This is 
straightforward for the red spectra since the acquisition images were taken with the same filter. For the blue spectra we have assumed 
$(V - R_C) \sim 1.8 \pm 0.2$ (as measured by McGehee et al. \cite{mcgehee} in February-April 2004 and from Kospal et al. \cite{kospal} 
in October 2005) from which we computed the absolute V flux level.  
\noindent

\begin{table*}
\caption{Log of spectroscopic observations with FORS2 of V1647 Ori.}
\label{tab:speclog}      
\centering           
\begin{tabular}{c c c c c c c c c c c }
\hline\hline           
Date       & JD -2 450 000   & FWHM & Slit width & Slit PA  &  Spectral range & Exposure time & S/N  & $R_C$ \\
 (UT)      &       & (\arcsec)  & (\arcsec)        &($^{\circ}$)& (\AA)  &(sec)      &   & (mag)\\
\hline
2004-02-18 & 3053.081 & 0.85 & 1.0 & 45.0 & 4560-5860 &  700     & 20   & 17.39 $\pm$  0.10\\ 
2004-02-23 & 3058.090 & 1.25 & 1.0 & 45.0 & 4560-5860 &  700     & 19   & 17.31 $\pm$  0.10\\
2004-03-13 & 3077.018 & 0.9  & 1.0 & 45.0 & 4560-5860 &  700     & 14   & 17.52 $\pm$  0.10\\
2004-03-18 & 3082.011 & 1.15 & 1.0 & 45.0 & 4560-5860 &  700     & 13   & 17.42 $\pm$  0.10\\
2004-03-27 & 3091.020 & 0.9  & 1.0 & 45.0 & 4560-5860 &  700     & 14   & 17.31 $\pm$  0.10\\
2004-12-08 & 3347.315 & 1.2  & 0.7 & 90.0 & 4560-5860 &  500     & 9    & 17.23 $\pm$  0.05\\
2004-12-21 & 3360.251 & 0.65 & 0.7 & 90.0 & 4560-5860 &  500     & 14   & 16.87 $\pm$  0.05\\
2005-01-05 & 3375.180 & 1.35 & 0.7 & 90.0 & 4560-5860 &  500     & 10   & 17.07 $\pm$  0.05\\
2005-02-18 & 3419.126 & 1.0  & 0.7 & 90.0 & 4560-5860 &  500     & 4    & 17.77 $\pm$  0.05\\
2005-02-29 & 3430.114 & 0.65 & 0.7 & 90.0 & 4560-5860 &  500     & 9    & 17.12 $\pm$  0.05\\
2005-03-15 & 3444.045 & 0.72 & 0.7 & 90.0 & 4560-5860 &  500     & 7    & 17.50 $\pm$  0.05\\
2005-12-27 & 3731.215 & 1.33 & 0.7 & 90.0 & 4560-5860 &  4 x 1800& 2    & 20.74 $\pm$  0.11\\
\hline	     									  	   
2004-12-08 & 3347.322 & 1.15 & 0.7 & 90.0 & 5750-7310 &  500     & 20   & 17.23 $\pm$  0.05\\
2004-12-21 & 3360.258 & 0.68 & 0.7 & 90.0 & 5750-7310 &  500     & 42   & 16.87 $\pm$  0.05\\
2005-01-05 & 3375.187 & 1.08 & 0.7 & 90.0 & 5750-7310 &  500     & 26   & 17.07 $\pm$  0.05\\
2005-02-18 & 3419.133 & 1.12 & 0.7 & 90.0 & 5750-7310 &  500     & 15   & 17.77 $\pm$  0.05\\
2005-02-29 & 3430.121 & 0.62 & 0.7 & 90.0 & 5750-7310 &  500     & 23   & 17.12 $\pm$  0.05\\
2005-03-15 & 3444.052 & 0.69 & 0.7 & 90.0 & 5750-7310 &  500     & 22   & 17.50 $\pm$  0.05\\
2006-01-29 & 3764.122 & 0.73 & 0.7 & 90.0 & 5750-7310 &  4 x 1800& 4    & 22.05 $\pm$  0.11\\
\hline\hline                                  
\end{tabular}
\end{table*}

Differential aperture photometry of V1647 Ori has been computed over an aperture radius of 2\farcs52 (10 pixels) from the acquisition 
images ($R_C$ filter). In paper I we computed the $R_C$ magnitude of two references stars in the FORS2 frames which were used to 
calibrate the instrumental magnitude of the acquisition images discussed here. The two stars were found to be not variable and to have 
the following $R_C$ magnitude: RA(J2000) = 05:46:09.71; DEC(J2000) = -00:03:31.1; $R_C$ = 20.08 $\pm$ 0.09 and RA(J2000) = 05:46:05.88; 
DEC(J2000) = -00:02:39.7; $R_C$ = 16.39 $\pm$ 0.02. The latter corresponds to the comparison star ``E'' in Semkov \cite{semkov5578} and 
\cite{semkov5683}, who measured $R_C$ = 16.39 $\pm$ 0.02. The results are listed in Table \ref{tab:speclog}.

\subsection{Mid-IR spectroscopy}
\subsubsection{TIMMI2}
Low-resolution ($\lambda$/$\Delta \lambda$ $\approx$ 200) $N^\prime$-band (7.7--13.0 $\mu$m) spectra of V1647 Ori (IRAS 05436$-$0007) 
were obtained on 2004 March 08 (JD 2453072.526) and 2006 January 10 (JD 2453746.663) using the TIMMI2 instrument on the ESO 3.6~m 
telescope at La Silla.  Sky subtraction was achieved by chopping in the North direction with an amplitude of 10\arcsec\, followed 
by a nodding pattern with the opposite direction and amplitude. The total integration time was 25 minutes per spectrum. The slit, with a 
width of 1\farcs2, was centered on a compact source detected in the $N^\prime$-band acquisition image. A spectrum of the reference star 
HD~37160 (K0III) was obtained before or after each IRAS 05436$-$0007 observation. Data were reduced using the usual steps of residual 
background subtraction, spectral extraction, and wavelength calibration. Correction for the telluric ozone absorption bands, as well as 
absolute flux calibration were achieved by ratio-ing the IRAS 05436$-$0007 spectrum to that of HD~37160, flux-calibrated using the 
spectral templates by Cohen et al. (\cite{cohen}). The resulting spectra are shown in Fig.~\ref{fig:timmi}. 
\subsubsection{VLTI/MIDI}
We also analyzed interferometric observations of V1647 Ori obtained with VLTI/MIDI on 3 nights between 2004 December 30 and 2005 
March 01. The interferometric data were previously analyzed by Abraham et al. (\cite{abraham06}). Hence we will discuss here only the MIDI 
low-resolution (R=30) spectrum acquired on each night by the instrument after the interferometric observation. 
\noindent

Using the MIA+EWS-1.3\footnote{{\it http://www.strw.leidenuniv.nl/~koehler/MIA+EWS-Manual}} software package, a fixed mask was applied to
the MIDI chopped spectrum. The background has been estimated from the off-source (sky) frames and then subtracted from the on-source 
frames. Finally the one-dimensional spectrum has been extracted. In the same way, spectra of MIDI calibrator stars were extracted.
\noindent

Aperture photometry was computed from the acquisition images with a narrow filter centered at 8.7 $\mu$m adopting a fixed aperture 
of 12 pixel (1\farcs0). The throughput of the two MIDI channels are, for an unknown reason, different, and the aperture photometry 
differs for the two telescopes. We used the results from the channel B which tend to be the more stable of the two. Conversion factors 
from counts to Jy were computed from the MIDI calibrators, whose 8.7 $\mu$m flux were evaluated from theoretical spectral energy 
distributions. Energy distributions have been derived from the Cohen list (\cite{cohen}) of infrared standard stars for the calibrator 
HD 37160 (F$_{8.7\mu m}$ = 11.5 Jy), and by matching the spectral type of the calibrator to stars in the Cohen list and scaling the 
spectrum with the ratio of their IRAS 12 micron fluxes for HD 107446 (F$_{8.7\mu m}$ = 37.9 Jy) and HD 50778 (F$_{8.7\mu m}$ = 28.9 Jy). 
The three calibrators were also used to compute the sensitivity function of MIDI (wavelength dependence of instrument's response).
The spectrum of V1647 Ori has been flux calibrated by multiplying it by the sensitivity function and by scaling it in order to match the 
8.7 $\mu$m flux. The result is shown in Figure~\ref{fig:timmi}.
\section{Results}\label{sec:results}
\subsection{Optical lightcurve}
\begin{figure*}
\begin{center}
\includegraphics[scale=0.55, angle=90]{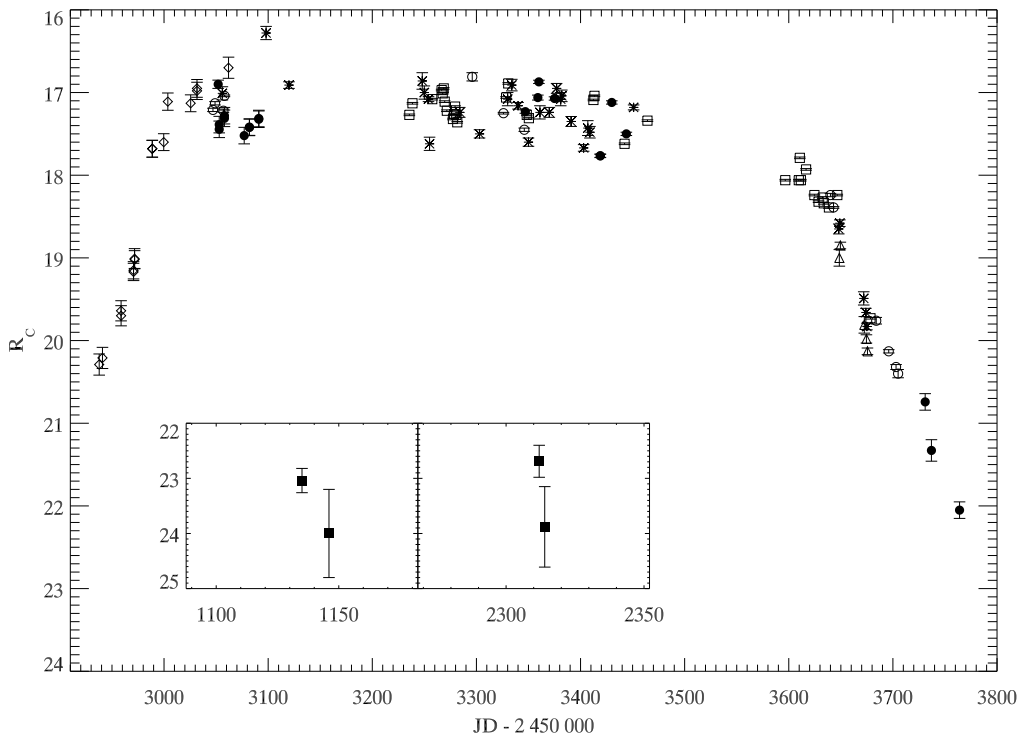}
\caption{Optical, $R_C$, light curve of V1647 Ori.
Data are from:  filled circles this work; squares Semkov, E.H. (\cite{semkov5683}, \cite{semkov5578}); triangles Kospal et al. 
(\cite{kospal}); stars \cite{acosta}; diamonds Brice\~no et al. (\cite{briceno}); filled squares McGehee et al. (\cite{mcgehee}); open 
circles Ojha et al. (\cite{ojha}); star \cite{acosta}. An offset of -0.3 mag was applied to the imported data (apart from those of Acosta-Pulido 
et al.2007) in order to reach a similar magnitude level with our data. In the inset the pre-outburst magnitude level reported by McGehee 
et al. (\cite{mcgehee}) is shown.}
\label{fig:lightcurve}
\end{center}
\end{figure*}

Figure~\ref{fig:lightcurve} shows the light curve of V1647 Ori in the $R_C$ band based on the data of Table~\ref{tab:speclog} and on
previous measurements by other authors. Due to the influence of McNeil's nebula on the computation of the stellar flux and subtraction 
of local background, measurements with different instruments may result in a different magnitude estimation. In particular, given the 
better spatial resolution of our data, we may better disentangle the contribution from V1647 Ori from that of the nebula. For this 
reason, an offset of -0.3 mag was applied to $R_C$ measurements by other authors. Such offset is not needed for the data of Acosta-Pulido
et al. 2007. Three further $R_C$ measurements from paper I are plotted in Figure~\ref{fig:lightcurve}.

Since measurements of $R_C$ of the early outburst are not available in literature, we estimated $R_C$ from I$_C$ measurements of 
Brice\~no et al. \cite{briceno}. The $(R_C - I_C)$ color, as 
measured by various authors (McGehee et al. \cite{mcgehee}, Ojha et al. \cite{ojha}), seems reasonably stable during the plateau and the 
fading phase, showing $(R_C - I_C) \approx$2.0 with a scatter of 0.2 mag. With this we calculate the expected $R_C$ magnitudes 
for the period October 2003 -- February 2004 from the $I_C$ measurements, assuming the same $(R_C - I_C)$ color for the rising phase. 
From figure \ref{fig:lightcurve}, the optical light curve of V1647 Ori can be divided in three parts: i) from October 2003 to February 
2004 -- the rising period; ii) from February 2004 to August 2005 -- the {\it plateau} phase and iii) from August 2005 to January 2006 -- 
the fading period. 
\noindent

The rising part is very steep: from October 2003 to January 2004 the optical magnitude increases by more than 3 magnitudes in $R_C$. From
the pre-outburst magnitude level, $R_C \sim 23.5$, computed by McGehee et al (see inset in figure \ref{fig:lightcurve}) it results that 
the total rise in brightness of V1647 Ori is $>$ 6 mag in $R_C$. From the light curve in figure \ref{fig:lightcurve} we find a rate of 
increase of $R_C$ of $\sim$ 1.5 mag$/$month. Assuming that this rate remained constant during all the rising phase, we estimate 
that the outburst began around the middle of August 2003, slightly earlier than October-November 2003 as found by Brice\~no et al. 
(\cite{briceno}).
\noindent

During the {\it plateau} phase the optical brightness shows a slow decline with time ($\Delta R_C = 0.02$mag/month), on top of which 
$R_C$ displays a non-periodic, flickered, oscillation on short timescale. The light curve in figure \ref{fig:lightcurve} combines data 
from different works which have used different instruments and aperture sizes. This may result in a systematic scatter of the data from 
work to work. However, an intrinsic variation of the optical brightness on short time scale is clearly present. From our data we measure
a variation of $\sim$ 0.5 mag between 2004 Feb. 17 ($R_C$ = 16.90 $\pm$ 0.05, paper I) and Feb. 18 ($R_C$ = 17.39 $\pm$ 0.10, 
Table~\ref{tab:speclog}). Thus, V1647 Ori at its maximum light shows an optical brightness variation on a time scale of 24 hours.
For five nights we have two consecutive acquisition images (separated by a few minutes) from which we searched for even smaller time 
scale variations of $R_C$, however, no significant change in optical brightness ($\Delta R_C > 0.1$ mag) are detectable from these 
measurements. The short timescale variability is of the same order of that found by other authors in optical and near infrared (Ojha et 
al. \cite{ojha}, Walter et al. \cite{walter}) on a timescale of a week. The total duration of the {\it plateau} phase is less than 2 years.
\noindent

Our photometry confirms the rapid brightness decrease announced by Kospal et al. (\cite{kospal}). Four months after the claimed onset 
of the fading phase (dated to August 2005, when Orion reappeared on the sky, however from the light curve it is clear that it started 
before), the brightness of V1647 Ori was still diminishing. From August 2005 to January 2006  $R_C$ dropped of 4 mag. On 2006 January 29,
the last $R_C$ measurements, we estimate $R_C = 22.05 \pm 0.11$ which is still more than 1 magnitude above the pre-outburst level of 
McGehee et al. (\cite{mcgehee}). From the light curve we estimate a fading rate of $\sim$ 0.8 mag$/$month during this phase. Assuming a 
constant fading rate, $R_C$ reached the pre-outburst level at the beginning of April 2006.

It is worth to note that the light curve is not symmetric, the rising and fading phase have a different slope. From the two different 
rates we may infer that the physical process which led to the beginning of the outburst is not simply the reverse of the process ending 
this outburst. The rate of brightness variation during the three phases are in agreement with that found by \cite{acosta}.

\subsection{Optical spectra}
\begin{figure*}
\centering
\includegraphics[width=6cm, angle=90]{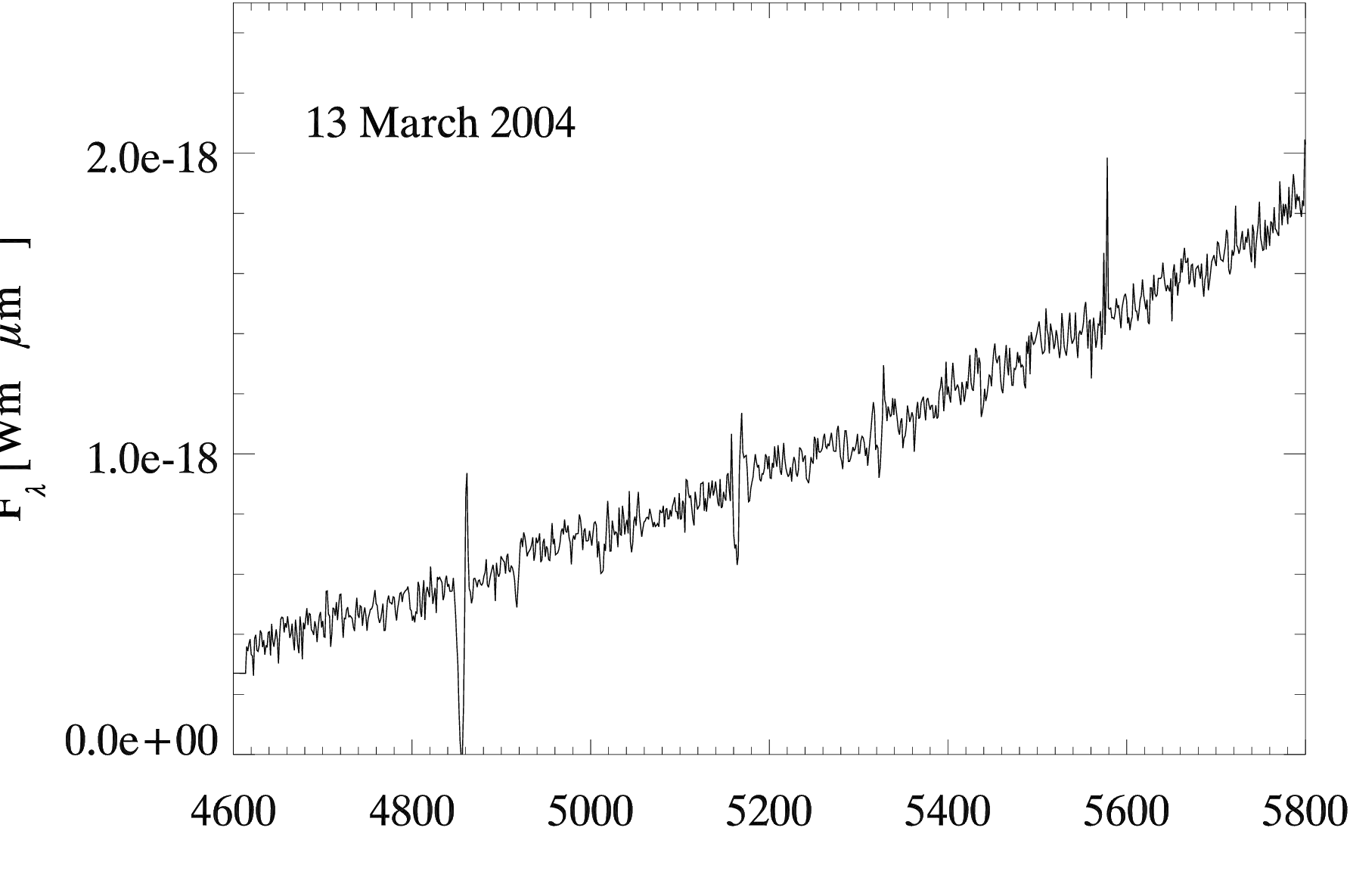}
\includegraphics[width=6cm, angle=90]{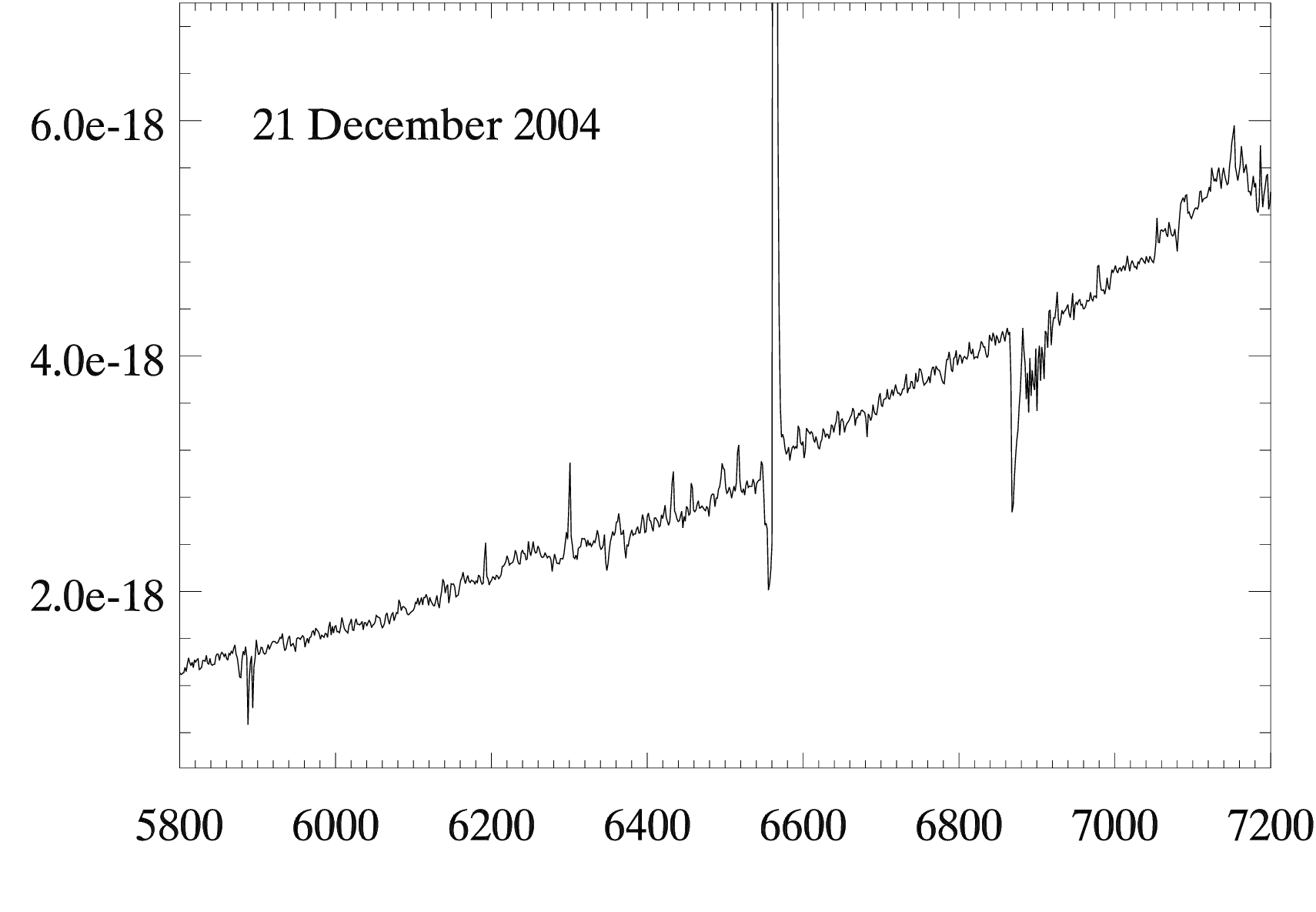}
\includegraphics[width=6cm, angle=90]{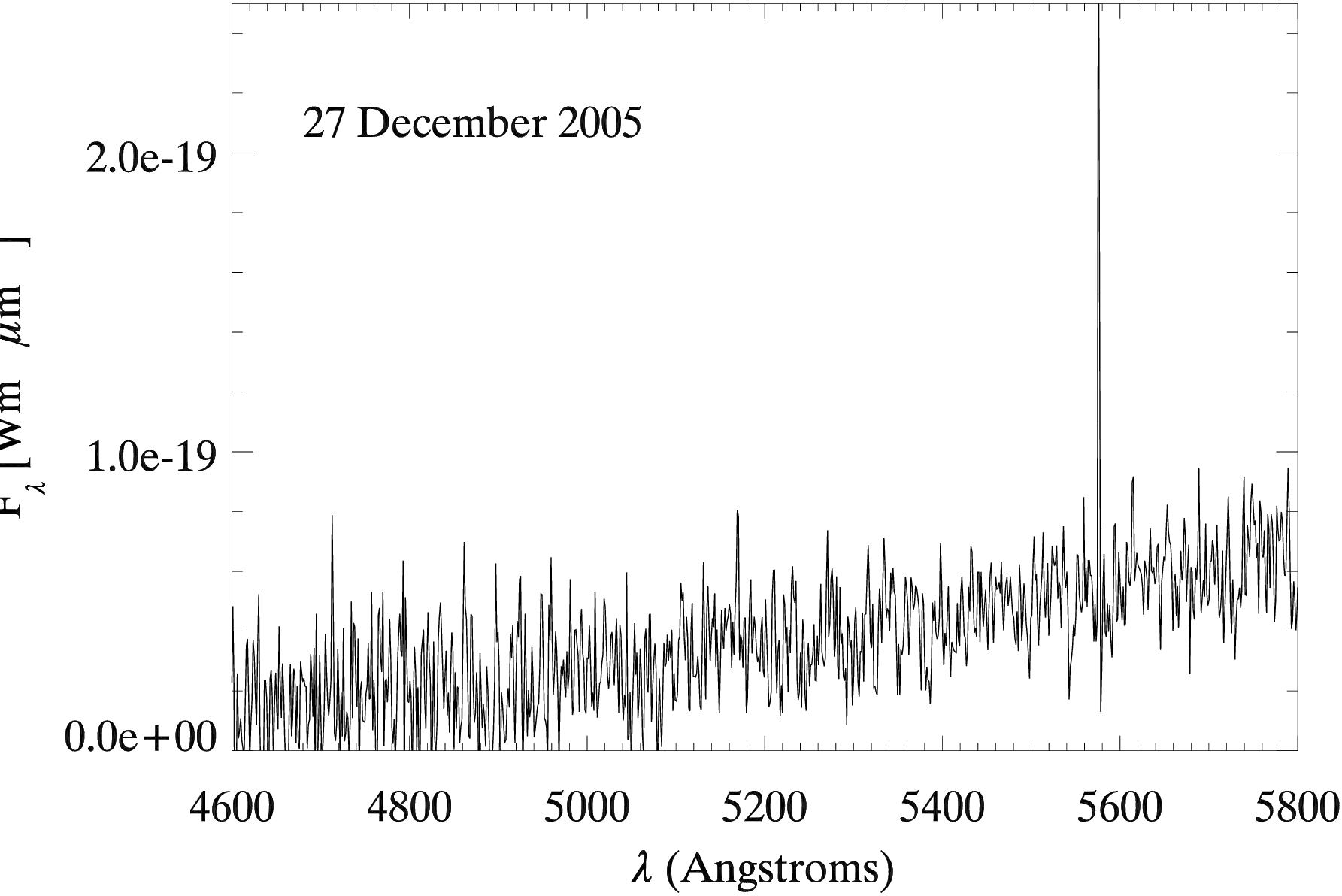}
\includegraphics[width=6cm, angle=90]{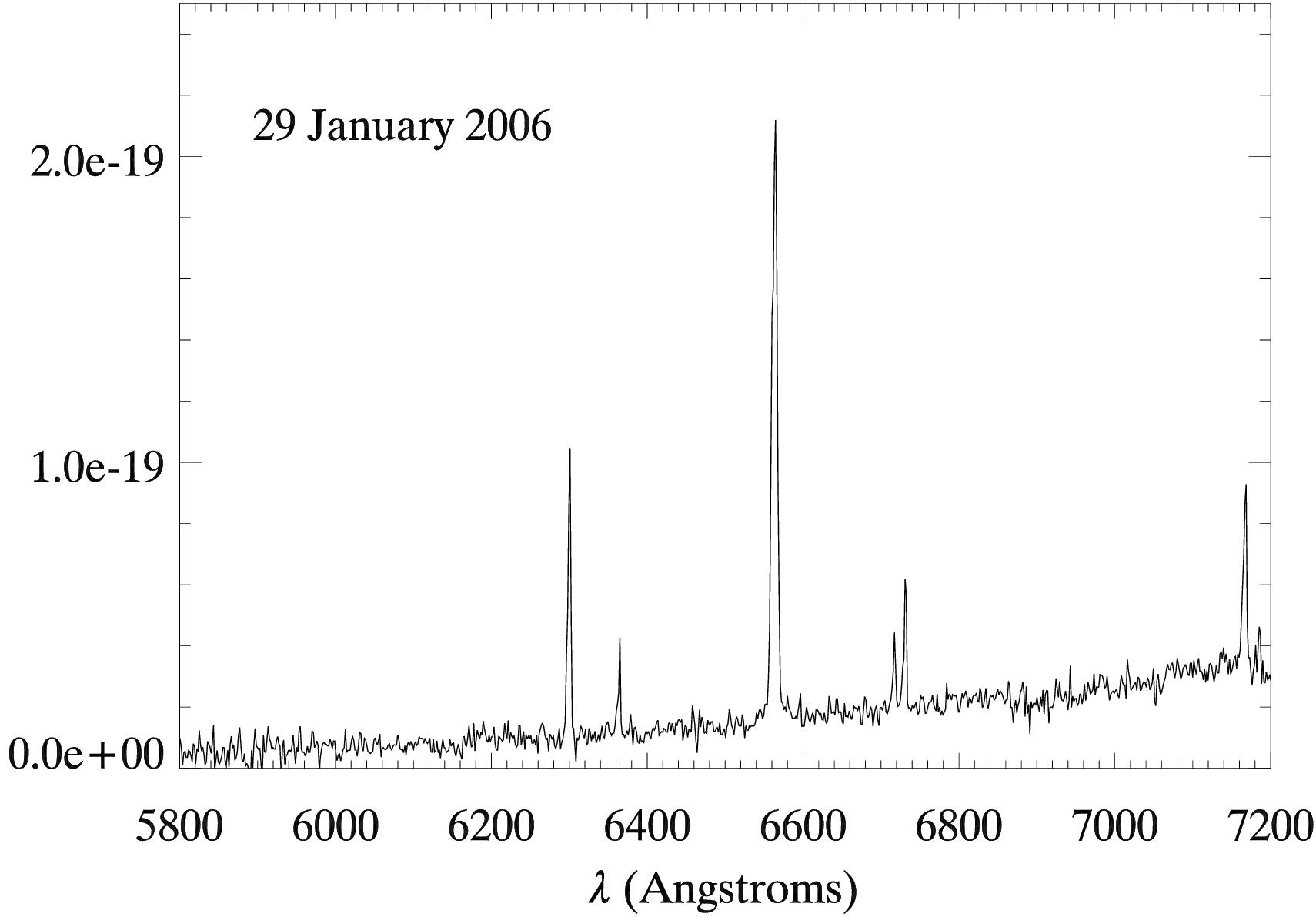}
\caption{Examples of optical spectrum of V1647 Ori obtained during the {\em plateau} phase (top two panels), and during the fading phase 
(bottom two panels). {\em Plateau spectrum}: clearly visible are the H$\alpha$ and H$\beta$ with P-Cygni profile, \ion{Fe}{I} 
(5328, 6191, 6495 \AA) and \ion{Fe}{II} (5169, 6432, 6516 \AA) lines and absorption from \ion{Na}{I} D1 \& D2 (5889, 5895 \AA) and 
\ion{He}{I} line (5875 \AA). A residual of the data reduction is present in correspondence of [\ion{O}{I}] (6300, 5577 \AA) due to bad pixel 
columns, and between 6870 - 6910 \AA ~due to a non-perfect removal of telluric absorption. {\em Fading phase spectrum}: no lines are 
detected in the blue part ([\ion{O}{I}] 5577 \AA ~is a residual of the data reduction), while the red part is characterized by a strong 
emission from H$\alpha$, [\ion{O}{I}] 6300, 6363 \AA, [\ion{S}{II}] 6717, 6731 \AA ~and [\ion{Fe}{II}] 7172 \AA ~lines.} \label{fig:spectra}
\end{figure*}

The positive slope of the optical spectrum of V1647 Ori (fig. \ref{fig:spectra}) reveals a red energy distribution of the source. Clearly
visible are the H$\alpha$ and H$\beta$ lines both characterized by a P-Cygni profile. The \ion{He}{I} $\lambda$5875, \ion{Na}{I} D1 \& D2
doublet in absorption and \ion{Fe}{I} and \ion{Fe}{II} lines in emission (P-Cygni) are detected in the spectra taking during the 
{\em plateau} phase. Due to bad pixel columns in correspondence of [\ion{O}{I}] $\lambda$5577 and [\ion{O}{1}] $\lambda$6300 lines a 
residual of the data reduction is present in the spectra taken during the {\em plateau} phase. For this reason we cannot find evidence of
any source features at these wavelengths.
\noindent

During the {\it plateau} phase the optical spectrum shows only minor changes on equivalent width and line flux (see discussion below). 
The overall shape of the spectrum remains ``constant'' during this period. The spectrum taken during the fading phase (blue -- 2005 
December 27, red -- 2006 January 29, Fig.~\ref{fig:spectra}) is completely different from all the others: superimposed on a very faint 
continuum with a slightly positive slope there are very strong emission lines in correspondence of H$\alpha$, 
[\ion{O}{I}] $\lambda\lambda$6300,6363, [\ion{S}{II}] $\lambda\lambda$6717,6731 and [\ion{Fe}{II}] at 7172 \AA~. The sky line 
[\ion{O}{I}] $\lambda$5577 is saturated and it is not possible to find evidence of source emission at this wavelength.
\noindent

From the flux-calibrated spectra, for each line detected, we computed the equivalent width (EW) and the line flux (F$_l$) multiplying 
EW by the continuum level at the line center (Tables \ref{tab:lines1} \& \ref{tab:lines2}). A major source of uncertainties in this 
computation is the determination of the continuum level, which in turn depends on the accuracy of the sensitivity function and of the 
aperture photometry from the acquisition images. The final accuracy is of the order of 10\% on EW and of 15\% on F$_l$.
\noindent

\begin{figure*}[!ht]
\centering
\includegraphics[angle=90, width=16cm]{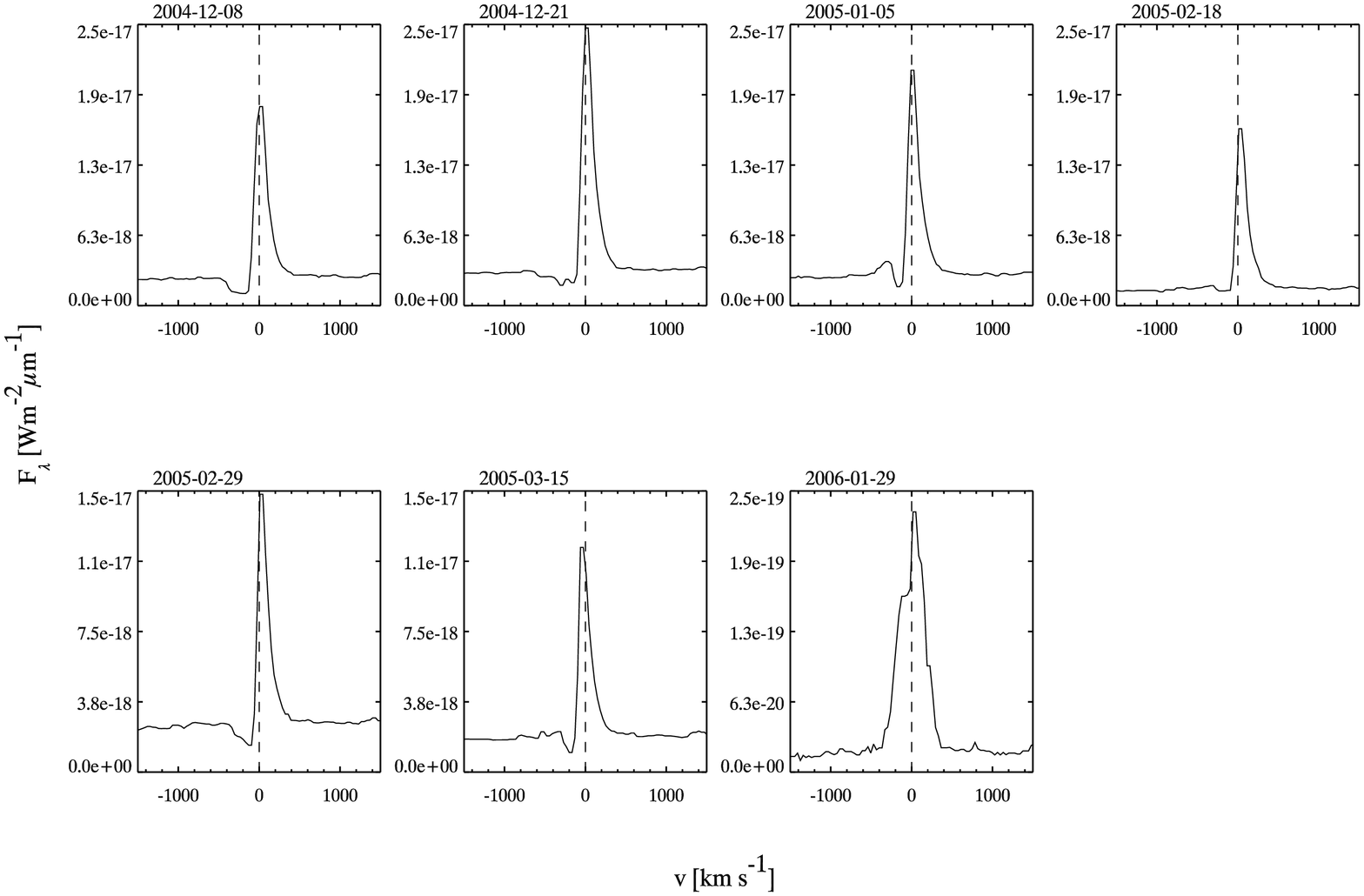}
\caption{Time evolution of the H$\alpha$ emission in the period December 2004 - January 2006. In all the ``{\em plateau} spectra'' the
line has an asymmetric P-Cygni profile. In the ``fading phase spectrum'', the line is pure in emission with no trace of absorption.}
\label{fig:ha}
\end{figure*}

\begin{figure*}[!ht]
\centering
\includegraphics[angle=90, width=16cm]{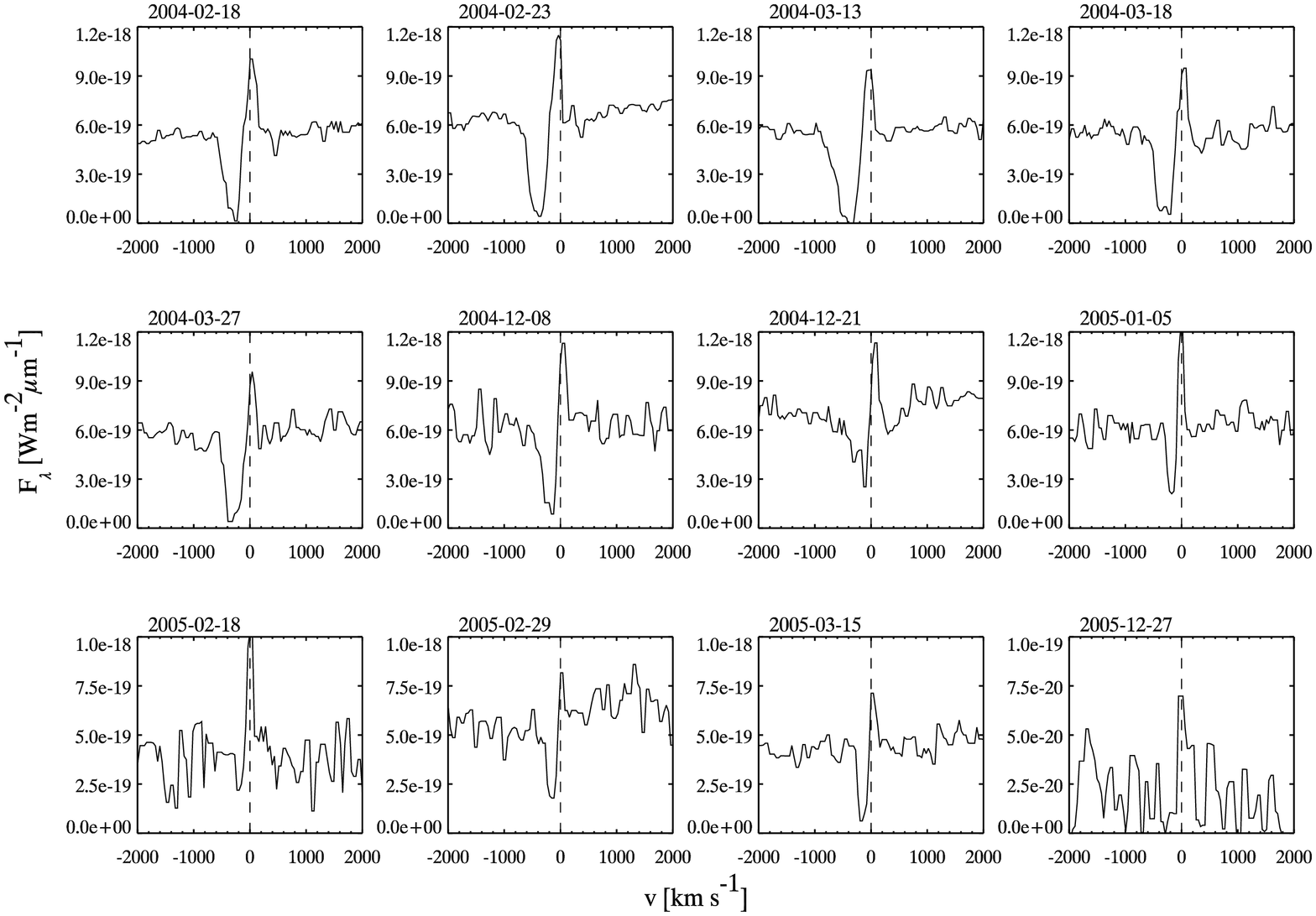}
\caption{Time evolution of the H$\beta$ emission in the period February 2004 - December 2005. As for the H$\alpha$, during the 
{\em plateau} phase the line has a P-Cygni profile. No H$\beta$ emission is detected in the low S/N ``fading phase spectrum''.}
\label{fig:hb}
\end{figure*}

The P-Cygni profile of H$\alpha$ and H$\beta$ is in both cases asymmetric with the emission components lacking the high velocity part 
(see Figures~\ref{fig:ha}, \ref{fig:hb}). This profile results from deviation of spherically-symmetric wind and is observed in FU Ori 
objects and T Tauri stars and can be explained by the presence of an opaque disk which occults part of the redshifted emission (see e.g. 
Hartmann \cite{hartmann}). The profiles of the two lines differs significantly (see figures \ref{fig:ha} \& \ref{fig:hb}): the H$\beta$ 
has a strong and wide absorption and a ``weak'' narrow emission while the H$\alpha$ has a huge emission and a weak absorption. In both 
cases, the blue-shifted absorption shows at least two components: one at $\sim$ --450 km s$^{-1}$, and the other at $\sim$ -- 150 km 
s$^{-1}$. While the low velocity component remains almost constant over all the {\it plateau} phase, the high velocity one is variable.
In particular, the latter shows in both lines a progressively decrease in extension from February 2004 to March 2005 until the whole 
absorption disappear in the fading phase spectrum. Furthermore, on three nights (2005 January 05, February 18 and March 15) the bluest 
absorption component of the H$\alpha$ is ``replaced'' by an emission. Also the emission component varies from night to night, displaying
a change in equivalent width and line flux.
\noindent


P Cygni signatures are also displayed by Fe lines. However, due to the low S/N of the spectra, the absorption component is clearly 
detected only for the \ion{Fe}{II} $\lambda$5169.08 transition. The maximum absorption is at $\sim$ -- 370 km s$^{-1}$ and has a terminal
velocity of $\sim$ -- 600 km s$^{-1}$. The other Fe emission lines show variation in equivalent width and line flux with time.
\noindent 

\begin{figure*}[!ht]
\centering
\includegraphics[width=8.5cm]{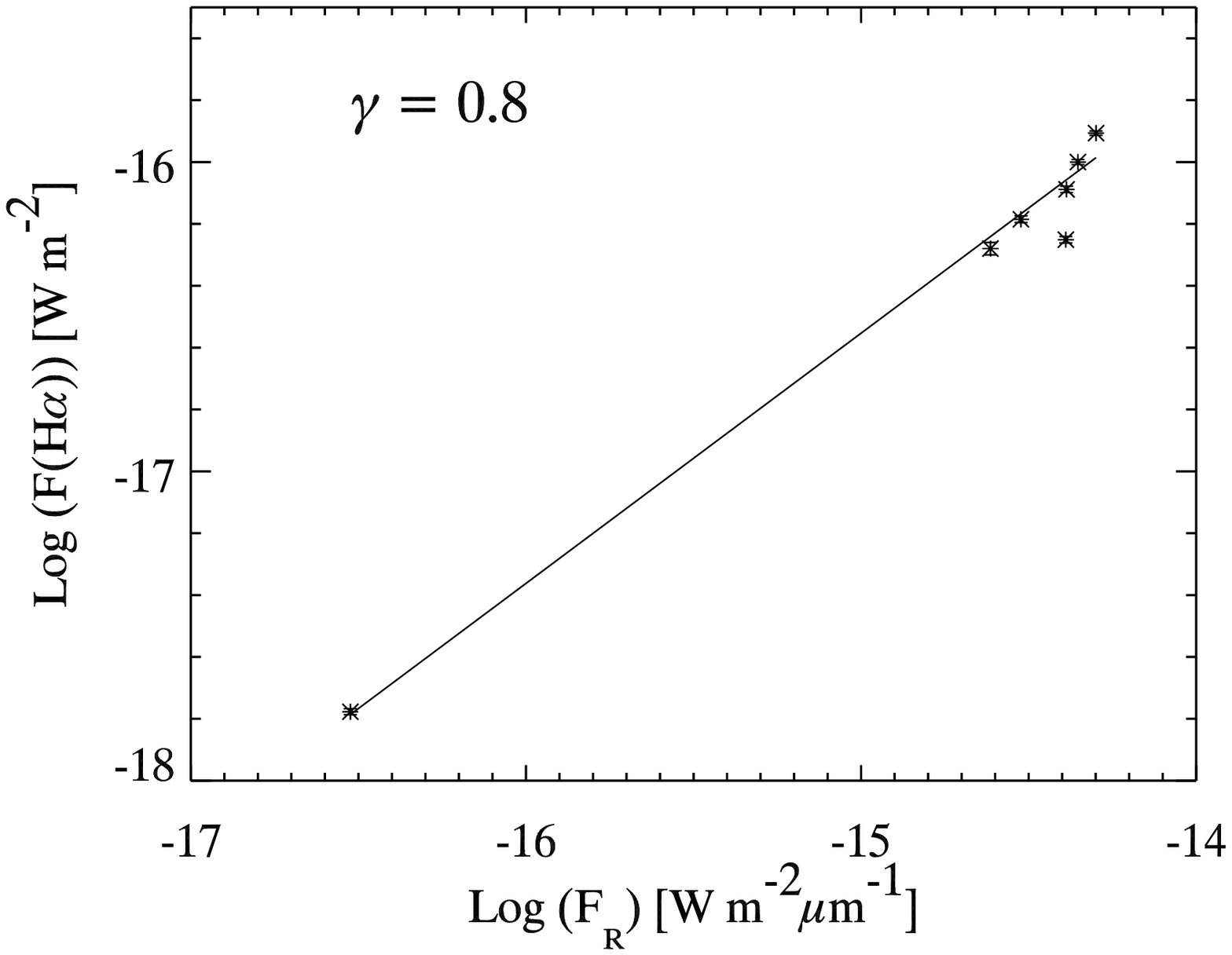}
\includegraphics[width=8.5cm]{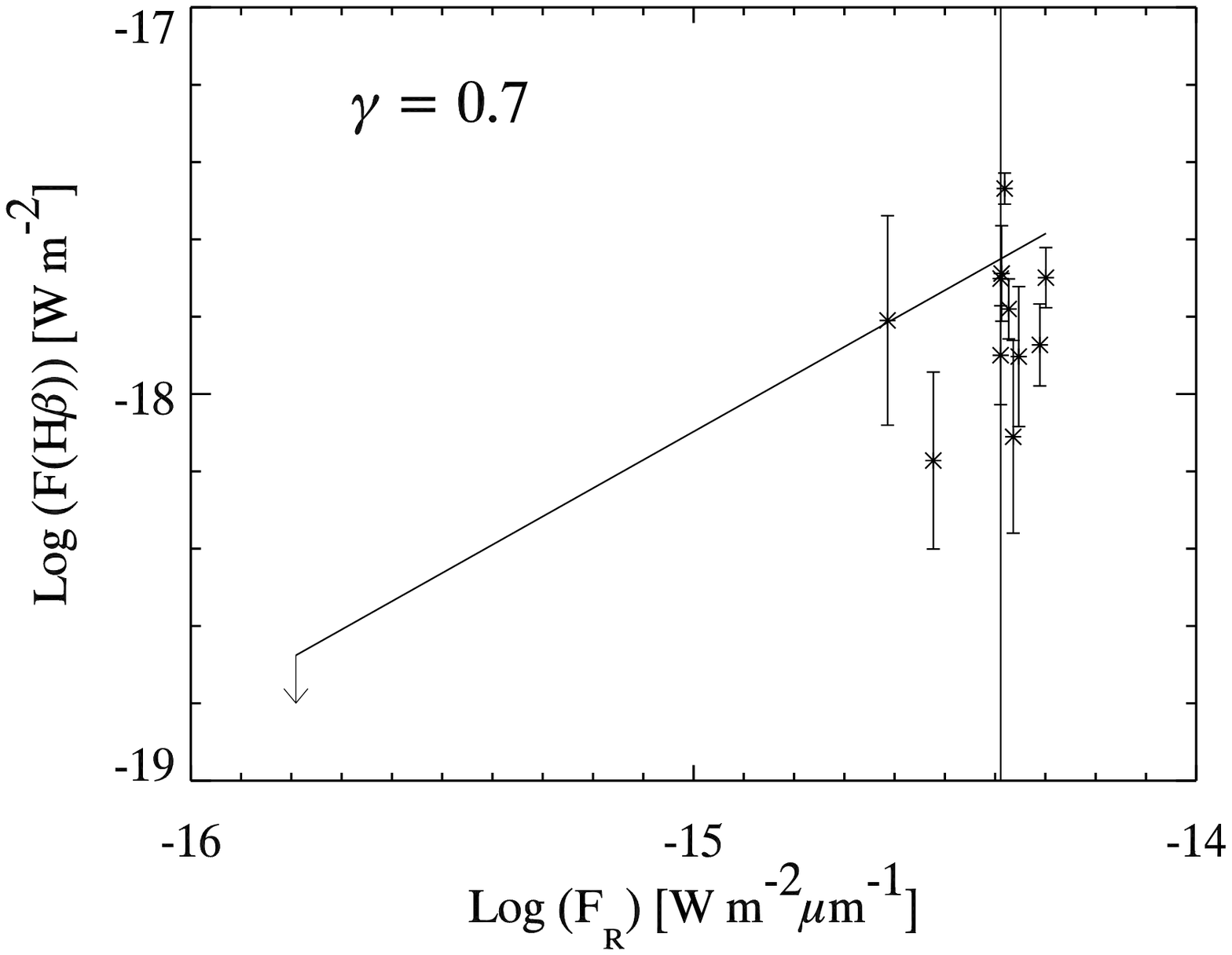}
\caption{Line flux of H$\alpha$ and H$\beta$ versus continuum flux. The best power law fit is over-plotted.}
\label{fig:hahb}
\end{figure*}

Thus, similar to what we have found for the continuum emission ($R_C$ magnitude), the emission lines vary on time scales of months and 
weeks. If continuum and line emission are produced in the same region, then the flux of a generic emission line ($F_l$) is linearly 
proportional to continuum flux ($F_R$ -- $R_C$ bandpass flux). In figures \ref{fig:hahb} we plotted the line flux of the strongest
emission lines detected (H$\alpha$ and H$\beta$) versus F$_R$. Arrows indicate upper limits. Assuming a power law dependence of the 
line emission from F$_R$ ($F_l \propto F_R^{\gamma}$) we searched for a correlation between line and continuum emission. The 
continuous line in figures \ref{fig:hahb} is the best fit to the data. We find $\gamma_{H\alpha} = 0.8$ and $\gamma_{H\beta} = 0.7$. 
The H$\alpha$ best fit is more robust than the H$\beta$ one (the H$\alpha$ emission is much stronger and it is also detected during 
the fading phase). All the other emission lines detected are weak and strongly affected by the low S/N of the spectra. The power law 
model fit in this case is more uncertain and clear evidence of dependence of the line emission from the continuum emission cannot be 
found. The value of $\gamma$ found for the H Balmer lines is close to unity. This means that the variations in line flux is 
correlated to the variation of the continuum emission. This evidence might suggest that the two emissions arise from the same region of 
the system, or at least, that the emission mechanisms are physically linked.
\noindent

Contrary to what was found by \cite{acosta}, we do not find any clear trend of the Balmer emission lines  with time during the 
{\em plateau} phase. Both the EW and the line flux show a random variation with time in this period.
\noindent

For the first time we detected optical forbidden lines in the spectrum of V1647 Ori taken during the fading phase (see figure 
\ref{fig:spectra}, bottom right). The spectrum clearly shows strong emission from [\ion{O}{I}] $\lambda\lambda$6300,6363, [\ion{S}{II}] 
$\lambda\lambda$6717,6731, [\ion{Fe}{II}] $\lambda$7172 an H$\alpha$. These lines provide evidence for hot (a few thousand K) gas close 
to V1647 Ori. These are used as tracers of Herbig-Haro objects where a collimated jet from the central star collides with the ambient 
medium. The emission is produced by the cooling of the shocked gas. It is interesting to note that Eisl\"offel \& Mundt 
(\cite{eisloffel}) have already identified IRAS05436-0007 as the driving source of HH 23. Similar and perhaps newly formed ejecta could 
be responsible for the forbidden emission lines seen here. None of these forbidden lines were previously detected in the {\it plateau} 
spectrum, most likely because of the overwhelming continuum.

\subsection{Mid-Infrared spectra}
\begin{figure}
\centering
\includegraphics[width=10cm]{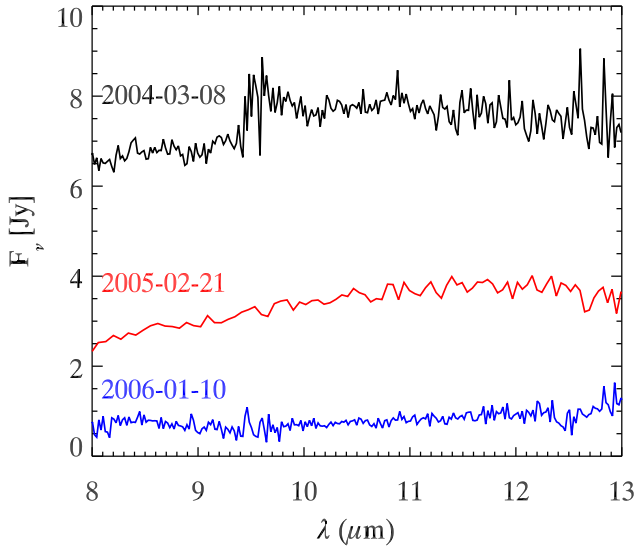}
\caption{Time evolution of the mid-Infrared spectrum of V1647 Ori taken with TIMMI2 (black and blue lines) and MIDI (red line).
The mid-infrared spectrum measured by MIDI remain constant, within the uncertainties (10\%), between December 2004 and March 2005. 
Here, only the MIDI spectrum taken on 2005 February 21 is shown.}\label{fig:timmi}
\end{figure}

\begin{table}
\caption{Mid-infrared photometry of V1647 Ori from TIMMI2 and MIDI data. For comparison in column 3 and 4 we report the corresponding 
$R_C$ magnitude and the $R_C$ - N1 color, where N1 is the magnitude at 8$\mu$m converted to the ESO mid infrared photometric system 
(\cite{vanderbliek}).
}
\label{tab:midir}
\centering
\begin{tabular}{lllll}
\hline\hline
DATE       & F$_{8\mu m}$  & $R_C$ & $R_C$ - $N1$  & Ref.   \\
           &   [Jy]        & [mag] & [mag]       &        \\
\hline	                               	    
2004-03-07 &   6.4         & 16.8  & 14.4        & Muzerolle et al.(\cite{muzerolle})\\
2004-03-08 &   6.5         & 16.8  & 14.4        & this work (TIMMI2)   \\
2004-03-11 &   4.3         & 16.8  & 14.0        & Andrews et al. (\cite{andrews})\\
2004-12-31 &   2.3         & 16.7  & 13.2        & this work (MIDI)     \\ 
2005-01-03 &   2.5         & 16.7  & 13.3        & this work (MIDI)     \\
2005-02-21 &   2.3         & 17.1  & 13.5        & this work (MIDI)     \\
2005-03-01 &   2.5         & 17.2  & 13.7        & this work (MIDI)     \\     
2006-01-10 &   0.5         & 21.0  & 15.8        & this work (TIMMI2)   \\
\hline
\hline
\end{tabular}
\end{table}
The rise in brightness of V1647 Ori during the outburst has been witnessed by others authors also at longer wavelength (see e.g. Andrews 
et al. \cite{andrews}, Muzerolle et al. \cite{muzerolle}, Abraham et al. \cite{abraham06}) . Our TIMMI2 spectra (see fig. 
\ref{fig:timmi}) confirm the increased mid-infrared flux: from the pre-outburst level of 0.53 Jy at 12 $\mu$m (IRAS, Point Source 
Catalog) up to 7.6 Jy on 2004 March 08. The $8 - 14 \mu$m spectrum is essentially featureless and flat all along the spectral range. This
result is not consistent with the mid-infrared spectrum in Andrews et al. (\cite{andrews}) taken with UKIRT/Michelle only three days 
after our TIMMI2 spectrum. Their measurement reveals a strong red energy distribution with the flux going from 4.0 Jy at 8 $\mu$m up to 
12.0 Jy at $12 \mu$m. Our estimate of the mid-infrared flux is however consistent with the SPITZER/IRAC observations of Muzerolle et al. 
(\cite{muzerolle}) taken on 2004 March 07. They measure indeed a flux of 6.43 Jy at 8 $\mu$m, very close to our estimate. A residual of 
the data reduction in correspondence of the strong atmospheric absorption bands centered at 9.6 (0$_3$) and 12.55 $\mu$m 
(CO$_2$) do not allow us to better analyze these two regions. 
\noindent

Ten months later, on December 2004, the mid-infrared flux of V1647 Ori revealed by the MIDI observations dropped by a few Jy. The 
spectrum is again flat and featureless. The MIDI spectra analyzed here, were all taken during the optical {\em plateau} phase 
(Fig.~\ref{fig:timmi}). They all reveal a flat and featureless spectrum. Within the accuracy of these spectra (10\%), the flux level 
remain constant between December 2004 and March 2005. Thus, also in the mid-IR the system experienced a {\em plateau} phase.
\noindent

The rapid optical brightness fading is also experienced by the system in the mid-infrared: on 2006 January 11, the flux level of our 
TIMMI2 spectrum at 12 $\mu$m is $0.9$ Jy, still considerably higher than the pre-outburst level. Also in this case the spectrum is flat 
and featureless.  
\noindent

\begin{figure}
\centering
\includegraphics[width=10cm]{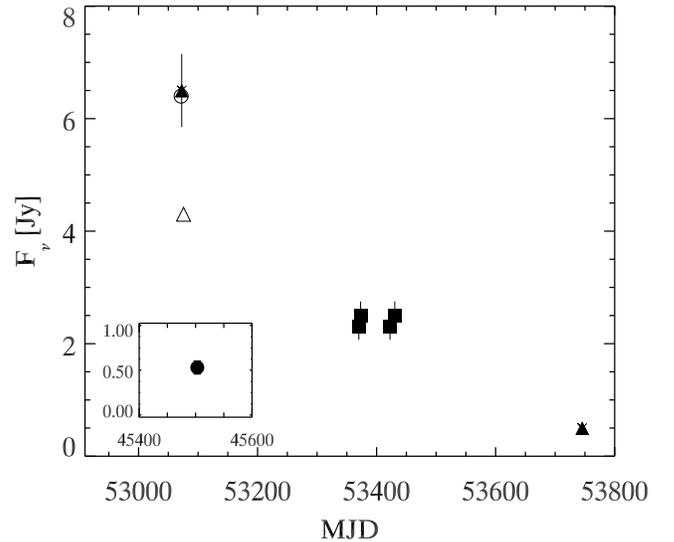}
\caption{8 $\mu$m light curve of V1647 Ori. Filled triangles - TIMMI2 data; filled squares - MIDI data; open circle - SPITZER/IRAC data 
from Muzerolle et al. (\cite{muzerolle}); open triangle - UKIRT/MICHELLE data from Andrews et al. (\cite{andrews}). The inset shows the 
pre-outburst IRAS measurements at 12 $\mu$m.}\label{fig:midIR-lc}
\end{figure}

It is worth to note that the optical and mid-IR light curve of V1647 Ori are different. In the optical the brightness increases during 
the rising phase remaining below the value of the {\em plateau} phase and finally decreases. On the other hand, the mid-IR brightness is 
higher during the ``mid-IR rising phase'' than during the ``mid-IR {\em plateau} phase'' (figure \ref{fig:midIR-lc}). 
\begin{table*}
\caption{Lines detected in the blue spectra of V1647 Ori between February 2004 and December 2005. EW is negative for emission lines and 
positive for absorption lines. In case of emission lines also the line flux is reported EW are expressed in \AA ~and line flux in 
10$^{-18}$W m$^{-2}$. The accuracy on equivalent width and line flux is respectively of the order of 10\% and 15\%.}\label{tab:lines1}
\centering           
\begin{tabular}{llllllllll}
\hline\hline            
Ident.         & $\lambda$[\AA] & \multicolumn{2}{c}{2004-02-18}& \multicolumn{2}{c}{2004-02-23} & \multicolumn{2}{c}{2004-03-13} & \multicolumn{2}{c}{2004-03-18}\\ 
               &                & EW      & F$_{line}$          & EW    & F$_{line}$             & EW    & F$_{line}$             & EW      & F$_{line}$\\
\hline																                                          
H$\beta$(em)   & 4861        &-2.91    &1.6                  &-1.94  &1.34                    &-5.38  &3.12                    &-2.22    &1.26      \\
H$\beta$(ab)   &             & 5.45    &                     & 5.91  &                        &11.48  &                        & 5.68    &          \\
$\ion{Fe}{II}$(em) & 5169    &-0.58    &0.51                 &-0.17  &0.18                    &-1.68  &1.52                    &-0.37    &0.32      \\
$\ion{Fe}{II}$(ab) &         & 0.74    &                     & 1.04  &                        & 2.16  &                        &0.98     &          \\
$\ion{Fe}{I}$  & 5328        &$>$-0.7  &$<$0.78              &-0.32  &0.39                    &-0.36  &0.39                    &$>$-0.6  & $<$0.7   \\
\hline\hline
Ident.         & $\lambda$[\AA] & \multicolumn{2}{c}{2004-03-27}& \multicolumn{2}{c}{2004-12-08} & \multicolumn{2}{c}{2004-12-21}   & \multicolumn{2}{c}{2005-01-05} \\ 
               &                & EW    & F$_{line}$            & EW    & F$_{line}$             & EW      & F$_{line}$             & EW     & F$_{line}$            \\
\hline							                                                                                      			         
H$\beta$(em)   & 4861        &-1.22   &0.75                  & -3.33     & 2.08               &-2.75    & 2.02      		   &-2.00   &1.27 \\
H$\beta$(ab)   &             & 5.53   &                      & 5.23      &                    & 6.05    &           	           & 2.15   &     \\
$\ion{Fe}{II}$(em) & 5169    &$>$-1.2 &$<$1.2                & -0.70     & 0.62               &-0.49    & 0.53        	   &$>$-1.3 &$<$1.2       \\
$\ion{Fe}{II}$(ab) &         & 1.34   &                      &$<$1.9     &                    &$<$1.2   &             	   &0.59    &             \\
$\ion{Fe}{I}$  & 5328        &-0.30   &0.34                  &$>$-1.1    &$<$1.2              &-0.29    &0.38         	   &-0.54   &0.62         \\  
\hline\hline            								            
Ident.         & $\lambda$[\AA] & \multicolumn{2}{c}{2005-02-18} & \multicolumn{2}{c}{2005-02-29}   & \multicolumn{2}{c}{2005-03-15} & \multicolumn{2}{c}{2005-12-27} \\
               &                & EW    & F$_{line}$             & EW      & F$_{line}$             & EW    & F$_{line}$             & EW    & F$_{line}$             \\ 
\hline			        				                                               							  
H$\beta$(em)   & 4861        &    -4.22  & 1.53               & $>$-1.5 &$<$0.94                 & -1.61      &0.70               &$>$-11.1 & $<$ 0.2  \\
H$\beta$(ab)   &                & $<$6.2    &                    &  2.55   &                        &  2.83      &                   &$<$22.1  &          \\
$\ion{Fe}{II}$(em) & 5169    & $>$-2.7   &$<$1.5              & $>$-1.2 &$<$1.1                  &$>$-1.1     &$<$0.8             &$>$-4.9  & $<$0.2   \\
$\ion{Fe}{II}$(ab) &            & $<$4.1    &                    & $<$1.9  &                        &$<$1.7      &                   &$<$ 7.3  &          \\
$\ion{Fe}{I}$  & 5328        & $>$-2.2   &$<$1.4              & $>$0.8  &$<$0.8                  &$>$-1.3     &$<$1.0             &$>$-4.5  & $<$0.2   \\
\hline\hline            								            
\end{tabular}
\end{table*}
\begin{table*}
\caption{Continuation of table \ref{tab:lines1} for red spectra between December 2004 and January 2006.}\label{tab:lines2} 
\centering           
\begin{tabular}{llllllllllllllll}
\hline\hline
Ident.&$\lambda$[\AA]         & \multicolumn{2}{c}{2004-12-08} & \multicolumn{2}{c}{2004-12-21} & \multicolumn{2}{c}{2005-01-05} & \multicolumn{2}{c}{2005-02-18}\\  
&                             & EW & F$_{line}$ & EW&F$_{line}$  & EW & F$_{line}$  & EW & F$_{line}$ \\      
\hline			                                                    
$\ion{He}{I}$      & 5875 &   1.14   &          & 0.71      &          &0.32       &           &$<$1.6    &         \\
$\ion{Na}{I}$ (D1) & 5889 &   1.44   &          & 0.97      &          &2.26       &           &   2.82   &         \\
$\ion{Na}{I}$ (D2) & 5895 &   1.01   &          & 0.70      &          &1.17       &           &   2.88   &         \\
$\ion{Fe}{I}$ (169)& 6191 &  -0.31   &   0.54   &-0.35      &-0.74     &-0.34      &0.63       &$>$0.08   &$<$0.8   \\
$[\ion{O}{I}]$     & 6300 &  --      & --       & --        & --       & --        &  --       &  --      &  --     \\
$[\ion{O}{I}]$     & 6363 &  --      & --       & --        & --       & --        &  --       &  --      &  --     \\
$\ion{Fe}{II}$     & 6432 &-0.74     &1.58      &-0.44      & 1.17     &-0.45      &1.03       &-0.48     &0.58     \\
$\ion{Fe}{I}$      & 6495 &-1.08     &2.46      &-0.55      &1.54      &-0.73      &1.76       &-0.83     &1.08     \\
$\ion{Fe}{II}$     & 6516 &-0.55     &1.27      &-0.43      &1.24      &-0.53      &1.30       &-0.69     &0.93     \\
H$\alpha$(em)      & 6562 &-26.61    &66.34     &-28.74     &88.66     &-30.27     &80.44      &-38.81    &56.60    \\
H$\alpha$(ab)      &      & 3.30     &          & 2.55      &          &  0.74     &           &$<$0.53   &         \\
$[\ion{S}{II}]$    & 6717 &  --      &  --      & --        & --       &  --       &  --       &  --      &  --     \\
$[\ion{S}{II}]$    & 6731 &  --      &  --      & --        & --       &  --       &  --       &  --      &  --     \\
$[\ion{Fe}{II}]$   & 7172 &  --      &  --      & --        & --       &  --       &  --       &  --      &  --     \\
\hline\hline                                  
Ident. & $\lambda$[\AA]       & \multicolumn{2}{c}{2005-02-29} & \multicolumn{2}{c}{2005-03-15} & \multicolumn{2}{c}{2006-01-29}\\ 
&                             & EW & F$_{line}$    & EW & F$_{line}$     & EW & F$_{line}$ & & \\    
\hline			                                                    
$\ion{He}{I}$      & 5875  &$<$0.8     &          &$<$1.1     &       &$<$10.9   &              \\
$\ion{Na}{I}$ (D1) & 5889  &   1.87    &          &   1.16    &       &$<$9.1    &              \\
$\ion{Na}{I}$ (D2) & 5895  &   0.87    &          &   1.51    &       &$<$7.3    &              \\
$\ion{Fe}{I}$ (169)& 6191  &-0.24      &0.44      &-0.30      &0.38   &$>$-3.3   &$<$0.03       \\
$[\ion{O}{I}]$     & 6300  & --        & --       &  --       &  --   &-40.98    &0.42          \\
$[\ion{O}{I}]$     & 6363  & --        & --       &  --       &  --   &-6.92     &0.08          \\
$\ion{Fe}{II}$     & 6432  &-0.42      &0.94      &-0.54      &0.85   &$>$-2.7   &$<$0.03       \\
$\ion{Fe}{I}$      & 6495  &-0.52      &1.23      &-0.92      &1.59   &$>$-2.7   &$<$0.03       \\
$\ion{Fe}{II}$     & 6516  &-0.35      &0.85      &-0.52      &0.92   &$>$-2.7   &$<$0.03       \\
H$\alpha$(em)      & 6562  &-17.25     &44.65     &-22.07     &39.94  &-108.30   &1.66          \\
H$\alpha$(ab)      &       & 1.78      &          & 1.10      &       &$<$3.9    &              \\
$[\ion{S}{II}]$    & 6717  & --        & --       &  --       &  --   &-4.52     &0.09          \\
$[\ion{S}{II}]$    & 6731  & --        & --       &  --       &  --   &-7.24     &0.15          \\
$[\ion{Fe}{II}]$   & 7172  & --        & --       &  --       &  --   &-9.48     &0.32          \\
\hline\hline
\end{tabular}
\end{table*}
\section{Discussion}\label{sec:discussion}
Pre-main-sequence stars are known to be intrinsically variable objects. The variability mechanisms might be different: solar-like coronal
activity, spots on the stellar surface, stellar pulsation, partial obscuration and subsequent clearing of the line of sight. These 
processes are however unable to generate the $\sim$ 44 $L_{\odot}$ luminosity increment produced by V1647 Ori and to produce the $\gtrsim$ 
6 optical magnitude burst lasting more than 2 years. To release such an amount of energy the existence of a secondary luminosity source 
is necessary. Similar brightening events from FU Orionis stars are explained by a sudden increase of the mass accretion rate from a 
circumstellar disk onto the central star. The increased accretion rate produces an accretion luminosity ($L_{acc} \propto \dot{M}$) which 
may overwhelm the stellar brightness. Such a process can explain both the dramatic brightening (from X-ray to the infrared) as well as the
strong H$\alpha$ emission observed in the recent outburst of V1647 Ori. Kastner et al. (\cite{kastner}) recently indeed confirmed that the
X-ray evolution of V1647 Ori in outburst reflects the near-infrared evolution and is consistent with the hypothesis of an increased mass 
accretion rate.
\noindent

As a consequence of the enhanced accretion rate a strong wind develop from the disk's surface. The blue-shifted absorption component 
of H$\alpha$ and H$\beta$ in the spectrum of V1647 Ori are signatures of this wind. The disappearance of the absorption component in the 
H$\alpha$ during the fading phase is a confirmation that the strong wind ceased and that the system has been going back to a phase of 
slow accretion. In this system the origin of the H Balmer emission lines is controversial since both the wind and the mass infall might 
contribute to the line formation. The magnetospheric accretion predicts a further contribution to the lines emissivity produced in the 
optically thin free-falling accretion columns.
\noindent

The \ion{Fe}{II} lines at 6432.68 and 6516.81 \AA~ detected in the spectra of V1647 Ori both originate from multiplet 40. They are 
commonly also seen in emission in the spectra of strongly accreting young stars such as V380 Ori (Rossi \cite{rossi}; Shevchenko 
\cite{shevchenko}), Z CMa (Hessman et al. \cite{hessman}; Garcia et al. \cite{garcia}; van den Ancker et al. \cite{vandenancker}), PV Cep
and MWC 1080 (Hernandez et al. \cite{hernandez}) and are amongst the strongest emission lines in such environments. These \ion{Fe}{II} 
lines appear to be absent from the spectrum of young stars showing lower accretion rates. The presence of \ion{Fe}{I} 6495.81 emission 
and the absence of strong \ion{Fe}{II} lines from higher multiplets suggests that the iron-line forming region in V1647 Ori may be 
somewhat cooler than the typical electron temperatures of 10,000-20,000 K inferred in for example Z CMa (van den Ancker et al. 
\cite{vandenancker}).

\subsection{The circumstellar envelope and the puzzling mid infrared spectrum}
The accretion disk alone is not able to produce the long wavelength ($\lambda \gtrsim$ 10 $\mu$m) emission observed, unless it flares 
strongly over a large range of distance scale (see e.g. Hartmann \cite{hartmann}). The sub-millimeter continuum flux during the outburst
remains at its pre-outburst level and there are no signatures of flux changes in these wavelength regime (Andrews et al. (\cite{andrews})).
These findings are consistent with the presence of a dusty circumstellar envelope, probably a remnant of the infalling envelope. 
Intriguingly, Kenyon \& Hartmann (\cite{kh}) suggested the presence of the infalling envelope to explain the enhanced mid-infrared flux
from FU Orionis objects in eruptive phase. 
\noindent
\begin{table*}
\caption{Typical value of outburst from pre-main-sequence stars. The outburst recurrence of FU Orionis objects has been estimated has the 
time needed to replenish the disk mass after an outburst with a constant infall rate.}\label{tab:fuors}
\centering
\begin{tabular}{|llll|}
\hline
                                                      & FUors             & EXors             & V1647 Ori     \\ 
\hline							          	                		   
Outburst duration               [yr]                  & $>$ 10            & $\sim$1           & 2.6           \\
Outburst recurrence             [yr]                  & $>$ 200           & 5 -- 10           & 37            \\
Mass accreted during an outburst[M$_\odot$]           & $>$ 10$^{-3}$     & 10$^{-6}-10^{-5}$ & 2.5$\cdot$10$^{-5}$ \\
Magnitude variation             [optical mag]         & 4 -- 6            & 2 -- 5            & $\sim$ 6      \\
Accretion Luminosity            [L$_{\odot}$]         & few 10$^2$        & $>$ 25            & 44            \\
Outburst accretion rate         [M$_\odot$ yr$^{-1}$] & 10$^{-4}$         & 10$^{-6}-10^{-5}$ & 10$^{-5}$     \\
Envelope infall rate            [M$_\odot$ yr$^{-1}$] & 5$\cdot$10$^{-6}$ & 10$^{-7}-10^{-6}$ & 7$\cdot$10$^{-7}$\\    
Wind velocity                   [Km s$^{-1}$]         & $>$300            & 200 -- 400        & 300 -- 400       \\
Mass loss rate                  [M$_\odot$ yr$^{-1}$] & 10$^{-6}-10^{-5}$ & 10$^{-8}-10^{-6}$ & 4$\cdot$10$^{-8}$ \\              
Spectral features & absorption spectrum     & emission line spectrum,            & emission line spectrum,                \\
                  & F/G-type supergiant like& T Tauri like, H$\alpha$ inverse P Cyg & strong H$\alpha$ emission (P Cygni) \\
                  & deep CO absorption      & CO abs./em., Br$\gamma$ emission & CO abs./em., Br$\gamma$ emission         \\
                  &                         &                                    & Forbidden lines in fading phase        \\
Note              &                         &                                    & X-rays variability                     \\
\hline
\end{tabular}
\end{table*}
Muzerolle et al. (\cite{muzerolle}, hereafter M04) attempt to reproduce the spectral energy distribution (SED) of V1647 Ori by means of a
standard viscous accretion disk and of an optically thin envelope. Their model predicts a 10$\mu$m emission feature that is produced by 
silicate dust grains. However, our multi-epoch mid-infrared spectroscopy reveals a flat and featureless spectrum during the whole 
outburst duration (see Fig. \ref{fig:timmi}).  This is highly unusual. In FU Orionis objects the silicate feature
is seen sometimes in emission (V1057 Cyg, FU Ori, BBW 76, V1515 Cyg, Green et al. \cite{green}) and sometimes in absorption (V346 Nor, 
Z CMA Green et al. \cite{green}, Acke \& van den Ancker \cite{acke}).  These differences are probably caused by differences in the 
optical thickness of the system (disk + envelope) at 10 $\mu$m.
\noindent
 
The model proposed by Abraham et al. (\cite{abraham06}) adopts a simple viscous accretion disk model (without envelope). Their model  
predicts a flat and featureless mid infrared spectrum. However, the emission at longer wavelength ($ \gtrsim 10 \mu m$) requires an 
highly flared disk. In such a model the outer part of the disk is directly illuminated during the outburst producing a flux enhancement
also in the (sub-)millimeter. This is not observed (Andrews et al. \cite{andrews}). The stability of the (sub-)millimeter emission favors
the presence of dusty circumstellar envelope.
\noindent

The excitation plot for the CO fundamental ro-vibrational lines (Rettig et al. \cite{rettig}) and the detection of $\Delta v=2$ CO 
band heads at $\sim$ 2.3 $\mu m$ (Vacca et al. {\cite{vacca}}) indicate the presence of hot ($\sim$ 2500 K) and dense gas. Since the 
dust sublimes at $\sim$ 1500 K, such emission likely arises in regions free of dust. Our suggestion is that even in the mid infrared the 
bulk of the emission is produced by the gas in a dust-free region of the disk. Nevertheless we cannot exclude a contribution from the 
dust (e.g arising at larger radii in the disk). The emission at longer wavelength is dominated by the dust in the envelope.
\noindent

Figure \ref{fig:midIR-lc} shows the temporal evolution of the mid infrared flux during the outburst. The flux decay at these wavelengths 
is faster than the decay in the optical (compare with figure \ref{fig:lightcurve}). The 8-14 $\mu$m flux drastically decreases from March
to December 2004 while the optical continuum remains constant over the same period. This produces radical changes in the SED of V1647 Ori
during the outburst. As can be seen from the $R_C$ - N1 color (Table \ref{tab:midir}) the system is redder during the early outburst. 
A likely explanation is that in the earlier phases of the outburst the disk was hotter and a larger region of the disk contributed to the
emission seen at 10 $\mu$m. However, a detailed modeling of the observations is necessary to explain which parameters (such as 
temperature, opacity, surface area) are responsible for the observed changes.
\section{Conclusion}\label{sec:conclusions}
Outbursts in pre-main-sequence stars have been historically classified in two main groups upon their similarity to the prototypes FU 
Orionis and EX Lupi (Herbig \cite{herbig77}) depending on outburst duration, maximum magnitude variation and spectral features at maximum
light. Table \ref{tab:fuors} lists the main characteristics of the two groups and that of V1647 Ori.
\noindent

Since the onset of the outburst of V1647 Ori it has been debated whether this system is either a FUor or an EXor object. V1647 Ori 
resembles some aspects of an EXor (outburst duration, recurrence of the outburst), and some aspects of a FUor (magnitude rise, SED). 
However the recurrence timescale of the outburst has intermediate value between the two classes. Its emission line spectrum is clearly 
distinct from either the absorption line spectrum of a FUor or the T Tauri-like spectrum of an EXor (where the H lines show inverse 
P-Cygni profile). 
\noindent

V1647 Orionis is not the only outburst PMSs suspected of having an intermediate nature between the two main classes. OO Ser experienced
recently an outburst which lasted $\sim$ 5-10 years (Kospal et al. \cite{kospal06}), too fast for a FUor and too slow for an EXor. The 
SED of OO Ser is typical of a FU Orionis object, and has roughly the same shape in quiescent and outburst phase. 
\noindent

A common denominator in all young eruptive stars detected so far seems to be the presence of circumstellar material as well as that of
a reflection nebula. These structures are likely remnants of the infalling envelope. The infalling envelope is a potential reservoir of 
mass for the disk which experiences recursive outbursts. Assuming that L$_{bol}$ during the outburst is dominated by the accretion 
luminosity, M04 estimate a mass accretion rate of $\sim$ 10$^{-5}$ M$_{\odot}$yr$^{-1}$. Considering the 2-3 years duration of the 
outburst and the 37 years recurrence timescale, a constant envelope infall rate of $\sim$ 7$\cdot$10$^{-7}$ M$_{\odot}$yr$^{-1}$ is 
necessary to replenish the disk after each outburst. The disk accretion rate during the quiescent phase is estimated to be $\sim$ 
6$\cdot$10$^{-7}$ M$_{\odot}$yr$^{-1}$ (see e.g. M04).
\noindent

Submillimeter maps reveal that FU Orionis stars have accretion disks that are larger and more massive than those of T Tauri stars 
(Sandell \& Weintraub \cite{sandell}) and are comparable in mass to those seen around Class I sources (i.e. young stellar objetcs with 
flat or rising infrared SED). The circumstellar material around V1647 Ori accounts for 0.04 $\pm$ 0.01 M$_{\odot}$ (Tsukagoshi et al. 
\cite{tsukagoshi}) which is slightly larger the disk mass of a T Tauri star ($\sim$ 0.01 M$_{\odot}$). All these findings suggest that 
outbursts occur in Class I sources, where the star is still embedded in the infalling envelope. The outburst duration and mass accretion 
rate during outburst seem to correlate with the infall rate (see Table \ref{tab:fuors}): objects with higher infall rate have longer 
outburst and reach higher accretion rate while objects with smaller infall rate experience short-lived outbursts. The occurrence of short
outbursts might suggest that the envelope is becoming more and more thin, i.e. that the system is in a transition phase of an embedded 
Class I source to an optically visible star surrounded by a protoplanetary disk (Class II).
\noindent 

The Orion Nebula Cluster (ONC) is subject to extensive observational campaigns. So far, roughly 1600 stars have been confirmed to be 
members of the ONC and 55\% of these (at least) posses a circumstellar disk (Hillenbrand et al. \cite{hillenbrand}). Assuming that Class I 
sources account for 20\% of stars with disks (lower limit, equal to the Class I fraction found in the original paper of Lada \& 
Wilking \cite{lada}), we expect to have more than 200 Class I sources in the ONC only. If all of these 200 sources experience FU 
Orionis-like activity, we would expect to see 10 -- 40 (depending on outburst duration and recurrence of FUors, V1647 Ori and EXors) of 
these stars in outburst at any time in the ONC. The total number of young eruptive stars discovered so far in the whole Orion star 
forming region is only seven: three FUors (FU Orionis itself, V883 Ori and Reipurth 50), three EXors (NY Ori, V1118 Ori, V1143 Ori) and 
V1647 Ori. Not all of these are in outburst at the same time. It appears that there is a deficit of observed outbursts in Orion. We
conclude that not all Class I sources undergo FU Orionis-like events for their entire lifetime. 
\noindent

A possible solution to this problem is that we are over-estimating the number of ``true'' Class I sources in Orion. The spectral energy 
distribution of an isolated T Tauri source seen with a disk close to edge-on may mimic a typical Class I SED. An alternative solution 
is that outbursts occur only in a specific stage of the early evolution, namely, in the transition phase of an embedded Class I source to
an optically visible T Tauri or Herbig AeBe star.
\noindent

We also caution that this result is based only on a small portion of Class I sources in Orion, namely, those not embedded in high 
density regions. Recent Spitzer observations show that there is a large number of such sources which lie in regions of extremely high 
extinction (Megeath et al. \cite{megeath}). An outburst in one of these objects could easily have been missed in the existing surveys of
Orion. More regular infrared surveys of star forming regions are badly needed to investigate the number of FUor-like outbursts in these
embedded sources.

\subsection{Outburst mechanism}
Instability mechanisms of different flavors have been proposed to explain the FU Orionis phenomenon. The gravitational forces of a 
companion star may perturb the disk enhancing accretion (see e.g. Bonnell \& Bastien \cite{bonnel}). Gravitational instability has been 
also proposed but, for this instability to occur the disk has to be cold and massive. The most accepted mechanism to trigger such 
outbursts is via thermal instability in the inner disk (see e.g. Lin \& Papaloizou \cite{lin}; Clarke et al. \cite{clarke90}; Kawazoe 
\& Mineshige \cite{kawazoe}; Bell et al. \cite{bell}). A key point of this model is that of an high accretion rate in the outer disk, of 
the order of few $\times$ 10$^{-6}$ M$_\odot$ yr$^{-1}$. According to this model, outbursts will occur as long as mass is deposited in 
the outer disk at such high rate. This implies that the outbursts will become shorter in time and smaller in amplitude as the infall into
the disk ceases. An alternative explanation for eruptive events has been suggested by Gammie (\cite{gammie}). He suggested that 
``dead zones'' of decreased accretion may develop in the case of magnetic viscosity. Material from the outer part of the disk may 
accumulate at the edge of such dead zones until high-accretion-rate episodes occur.
\noindent

One way to explain the different outburst properties of FUors, EXors and V1647 Ori is by introducing a different outburst mechanism
for each of these classes of objects. However, the data collected so far in all young eruptive stars raise the possibility of the existence
of only a unique class of outburst objects. In this scenario, the different types of outburst are produced by a continuum variation of
one or more of the parameters involved in the instability, rather than a variation of physical mechanisms. For instance, the 
presence of a remnant of the infalling envelope and the estimated infall rate (see Table \ref{tab:fuors}) might favor the thermal 
instability as the unique model to explain the three groups. If this is the case, the different outburst duration between a FU Orionis,
V1647 Ori and an EX Lupi might be explained with the difference in the infall rate. 
The thermal instability model predicts indeed shorter and smaller amplitude outbursts as infall ceases below 10$^{-6}$ M$_\odot$ yr$^{-1}$
to approach typical T Tauri disk accretion rate of $\sim$ 10$^{-7}$ M$_\odot$ yr$^{-1}$.
\noindent

The same model is able to explain also the diversity in spectral features of the three classes of objects. The mass accretion rate reached 
during the outburst is (also) dependent on the envelope infall rate. When very high accretion rates and disk temperatures are reached, as in 
the case of FUors, the disk's interior may become hotter than the disk surface and the emitted spectrum will show absorption lines. If the
accretion rate is not high enough to invert the temperature gradient in the disk's interior, the final spectrum will be dominated by emission
features which are produced in the disk's wind and/or in the magnetospheric accretion columns. In this case we will see EXors or 
V1647 Ori-like spectra.
\noindent

\begin{acknowledgements}
The authors thank the ESO Paranal and La Silla staff for performing the service mode observations. 
\end{acknowledgements}

%%%%%%%%%%%%%%%%%%%%%%%%%%%%%%%%%%%%%%%%%%%%%%%%%%%%%%%%%%%%%%%%%%%%%%%%%%%%%%%%%%%%%%%%%%%%%%%%%%%%%%%%
%%%%%%%%%%%%%%%%%%%%%%%%%%%%%%%%%%%%%%%%%%%%%%%%%%%%%%%%%%%%%%%%%%%%%%%%%%%%%%%%%%%%%%%%%%%%%%%%%%%%%%%%%%%%%%%%%%%%%%%%%%%
%

\begin{thebibliography}{99}
\bibitem[2004]{abraham04}
  Abraham P., Kospal A., Csizmadia A. et al. 2004, A\&A, 419,L39
\bibitem[2006]{abraham06}
  Abraham, P., Mosoni, L., Henning, T. et al. 2006, A\&A, 449,L13
\bibitem[2004]{acke} 
  Acke, B., \& van den Ancker, M.~E.\ 2004, \aap, 426, 151 
\bibitem[Acosta-Pulido et al. (2007)]{acosta} 
  Acosta-Pulido, J.~A., et al.\ 2007, astro-ph/0408432 
\bibitem[2004]{andrews}
  Andrews, S.M., Rothberg, B., Simon, T. 2004, \apj, 610,L45
\bibitem[Appenzeller et al. 1998]{appenzeller}
  Appenzeller, I. et al. 1998, The Messenger ,94,1 
\bibitem[2006]{aspin} 
  Aspin, C., Barbieri, C., Boschi, F., et al. \ 2006, \aj, 132,1298
\bibitem[1995]{bell}
  Bell, K.R., et al. 1995, \apj, 444,376
\bibitem[1992]{bonnel}
  Bonnel, I. \& Bastien, P. 1992, \apj, 401, 654
\bibitem[2004]{briceno} 
  Brice\~no, C., Vivas, A.K., Hernandez, J. et al. 2004, \apj, 606, L123
\bibitem[1990]{clarke90}
  Clarke, C., Lin, B.D.C.,\& Pringle, J.E., 1990 \mnras, 242, 439
\bibitem[2005]{clarke}
  Clarke, C., et al. 2005 \mnras, 361, 942
\bibitem[Cohen et al. 1999]{cohen} 
  Cohen, M., et al.  1999, \aj, 117,1864
\bibitem[1997]{eisloffel}
  Eisl\"offel, J., \& Mundt, R. 1997, \aj, 114, 280
\bibitem[2007]{fedele}
  Fedele, D. , van den Ancker, M.E., Petr-Gotzens, M.G. et al. 2007 submitted
\bibitem[1996]{gammie}
  Gammie, C.F. 1996, \apj, 457, 355
\bibitem[1999]{garcia}
  Garcia et al. 1999, A\&A, 346, 892
\bibitem[2006]{green}
  Green, J.D. et al. 2006 \apj, 648,1099 
\bibitem[2005]{grosso}
  Grosso, N., Kastner, J.H., Ozawa, H. et al. 2005 A\&A 438, 159
\bibitem[1996]{hk}
  Hartmann, L. \& Kenyon S.J. 1996 ARA\&A 34, 207 
\bibitem[1998]{hartmann}
  Hartmann, L., 1998 {\it Accretion processes in star formation}
\bibitem[1966]{herbig66}
  Herbig, G.H. 1966 Vistas in Astronomy 8, 10
\bibitem[1977]{herbig77}
  Herbig, G.H. 1977 \apj, 217, 693
\bibitem[2004]{hernandez}
  Hernandez et al. 2004, \aj, 127, 1682
\bibitem[1991]{hessman}
  Hessman et al. 1991, \apj, 370, 384
\bibitem[1998]{hillenbrand}
  Hillenbrand et al. 1998, AJ 116, 1816
\bibitem[2006]{kastner}
  Kastner, J.~H., et al.\ 2006, \apjl, 648,L43 
\bibitem[1993]{kawazoe}
  Kawazoe, E. \& Mineshige, S. 1993 \pasp, 45, 715
\bibitem[K\"aufl et al. 2003]{kaeufl}
  K\"aufl, H-U. et al. 2003, SPIE 4841, 117
\bibitem[2000]{kenyon}
  Kenyon S.J., et al. 2000 \apj, 531, 1028
\bibitem[1991]{kh}
  Kenyon, S.J., Hartmann, L.W. 1991 \apj, ~383, 664
\bibitem[2005]{kospal} 
  Kospal, A., Abraham, P., Acosta-Pulido, J., 2005 IBVS, 5661.
\bibitem[2006]{kospal06} 
  Kospal, A., Abraham, P., Prusti, T., et al. 2006 ASPC, 349, 269 
\bibitem[1984]{lada}
  Lada \& Wilking 1984, \apj, 287, 610
\bibitem[2003]{leinert}
  Leinert, Ch., Graser, U., Przygodda, F. et al. 2003, Ap\&SS, 285, 73
\bibitem[1985]{lin}
  Lin, B.D.C., \& Papaloizou, J.C.B. 1985 {\em Protostars and Planets II} 
\bibitem[1999]{magnier}
  Magnier E.A., Waters L.B.F.M., Groot P.J. et al. 1999 A\&A, 346, 441
\bibitem[Mallas \& Kreimer (1970)]{mallas}
  Mallas, J.H., \& Kreimer, E. 1970 {\em The Messier Album} 
\bibitem[2004]{mcgehee} 
  McGehee, P.M., Smith, J.A., Henden, A.A. et al. 2004 \apj, 616, 1058
\bibitem[2004]{mcneil}
  McNeil, J.W. 2004 IAU Circ. 8284
\bibitem[2006]{megeath}
  Megeath, S.T., Allen, L.E., Allgaier, E. et al. 2006 IAUS, 237, 167 
\bibitem[2005]{muzerolle}
  Muzerolle, J., et al. 2005 \apj, 620, L107
\bibitem[2006]{ojha}
  Ojha, D. K., Ghosh, S. K., Tej, A. et al. 2006 \mnras, 368, 825
\bibitem[2004]{reipurth}
  Reipurth, B., Aspin, C., 2004 \apj, 606, L119
\bibitem[2005]{rettig}
  Rettig, T.W., Brittain, S.D., Gibb, E.L. et al. 2005 \apj, 626,245 
\bibitem[1999]{rossi}
  Rossi, et al. 1999, A\&AS, 136, 95
\bibitem[2001]{sandell}
  Sandell, G. \& Weintraub, D.A. 2001 \apj, 134, 115
\bibitem[1999]{shevchenko}
  Shevchenko, 1999, Astronomy Reports 43, 246
\bibitem[2004]{semkov5578} 
  Semkov, E.H. 2004 IBVS, 5578.
\bibitem[2006]{semkov5683} 
  Semkov, E.H. 2004 IBVS, 5683.
\bibitem[2002]{smith}
  Smith J., et al. 2002 \aj, ~123, 2121
\bibitem[2000]{stetson}
  Stetson P., 2000 \pasp, ~112, 925
\bibitem[2005]{tsukagoshi}
  Tsukagoshi, T. et al. 2005 PASJ, 57, L21
\bibitem[2004]{vacca}
  Vacca, W.D., Cushing, M.C. Simon, T. 2004 \apj, 609, L29
\bibitem[2004]{vandenancker}
  Van den Ancker et al. 2004, MNRAS, 349, 1516
\bibitem[Van der Bliek et al. 1996]{vanderbliek}
  Van der Bliek, N.S., et al. 1996 A\&AS, 109,547
\bibitem[2004]{walter} 
  Walter, F.~M., et al. \ 2004, \aj, 128, 1872 
\end{thebibliography}
\end{document}